\documentclass[prl,twocolumn,preprintnumbers,amsmath,amssymb]{revtex4}

\usepackage{graphicx}
\usepackage{dcolumn}
\usepackage{bm}

\begin{document}


\title{Williams et al. Reply to the Comment by Dumin on \\``Progress in Lunar Laser Ranging Tests of Relativistic Gravity'' }

\author{James G. Williams}
\author{Slava G. Turyshev} 
\author{Dale H. Boggs} 
\affiliation{
Jet Propulsion Laboratory, California Institute of  Technology,
Pasadena, CA 91109, USA
}

\date{\today}

\begin{abstract}
A decreasing gravitational constant, $G$, coupled with angular momentum conservation is expected to increrase a planetary semimajor axis, $a$, as $\dot a/a=-\dot G/G$. Analysis of lunar laser ranging data strongly limits such temporal variations and constrains a local ($\sim$1~AU) scale expansion of the solar system as $\dot a/a=-\dot G/G =-(4\pm9)\times10^{-13}$~yr$^{-1}$, 
including that due to cosmological effects.

\end{abstract}

%
\maketitle


A connection of the cosmological expansion of the universe to processes on  smaller scales is an important problem which has been extensively studied in the past, but still awaits a proper solution (see \cite{cosmology2body} and references  therein).  In the weak-field and slow-motion limit of a metric that interpolates between a static Schwarzschild metric at small distances from a point mass, $M$, and a time dependent Friedmann spacetime at large distances \cite{cosmology2body,cosmology-solar-system}, the corresponding effects on the local dynamics are best exemplified by the corrections induced in the two-body problem:
$
\ddot {\vec r} - ({\ddot s}/{s}){\vec r} =- {GM}{\vec r}/{r^3},
$
with $s$ being the cosmological scale factor. Although similar results can be derived with Eq.~(1) of \cite{dumin}, the Comment does not make this connection. Instead, it implies that effects of $({\ddot s}/{s})$ may be seen as local expansion of the scale of our solar system ($\sim1$~AU); however, local scale effects were found to be many orders of magnitude too small to be observed \cite{cosmology2body,cosmology-solar-system}, making the expansion claim unjustified.  

Indirectly, cosmology may still affect the dynamics on local scales. Specifically, the cosmological evolution may cause variation of the effective 4-dimentional gravitational constant via its hypothetical dependence on extra space-time dimensions and/or new fields of matter \cite{G-dot-theory}. Our Letter \cite{LLR_beta_2004} reported on the investigation of such a scenario via the lunar laser ranging (LLR) experiment. 

The LLR data analysis uses an $n$-body relativistic numerical integration of the orbits of the Moon and planets \cite{Williams-etal-1996}. The least-squares fit for solution parameters uses a numerically integrated $\dot G$ partial derivative based on those same relativistic equations of motion.  It is convenient to compare the  $\dot G$ result with the Hubble constant using ${\dot G}/G=\sigma H$; at a minimum it makes an interesting comparison and a deeper connection is a possibility \cite{G-dot-theory}.   

The LLR $\dot G$ result is pertinent to any source of solar system expansion which conserves angular momentum. As discussed in \cite{Williams-etal-1996}, the stronger part of the LLR $\dot G$ determination comes from the accelerated angular motion of the Earth-Moon system about the Sun, rather than the linear change in orbit radius. Based on $\dot n/n=2\dot G/G$, with $n$ being the mean motion of an orbit, the angular motion behaves as $\Delta\varphi \propto t^2$; in addition, using $\dot a/a=-\dot G/G$, the relevant effect of changing $a$ is given by $\Delta a\propto t$. With a span of LLR data of more than three decades, the $t^2$ sensitivity is roughly two orders of magnitude larger than the linear term. 
Thus, radial expansion coupled with angular momentum  conservation is expected to give a nonzero $\dot G$ result whether or not $\dot G$ is the cause
($a\Delta\varphi\gg \Delta a$), but radial changes are excluded if they do not affect orbit period (angular motion).

While a change in the orbit period provides more sensitivity ($a\Delta\varphi$) than a radial change ($\Delta a$), a radial change without period change was tested in \cite{Williams-etal-1996}. An anomalous rate of change of the lunar mean distance was excluded if more than 3.5 mm/yr or $0.9\times 10^{-11}$~yr$^{-1}$ relative which is eight times smaller than the inverse Hubble time.

Thus, radial and angular changes are both excluded at values considerably less than the Hubble constant. Other than these changes, how would local  expansion affect the dynamics? To answer this question, one needs a description of the LLR observables in a metric that links local and cosmic scales, which is missing in the Comment.  

The value $\dot G/G=(4\pm9)\times10^{-13}$ yr$^{-1}$ reported in \cite{LLR_beta_2004} is the most accurate published limit using LLR, or $\dot a/a=-(4\pm9)\times10^{-13}$~yr$^{-1}$, implying a small radial drift of $-(6\pm13)$ cm/yr in an orbit at 1~AU.

Echoing the statement in \cite{LLR_beta_2004}, LLR data analysis does not support local ($\sim$1 AU) scale expansion of the solar system at the Hubble rate.  
If expansion exists, the effects on the solar system 
are likely very small; we encourage the search for a metric which describes the influence of the cosmic expansion locally. We also advocate solar system tests of such theoretical developments.

The work described here was carried out at the Jet Propulsion Laboratory, California Institute of Technology, under a contract with the National Aeronautics and Space Administration.



\end{document}